\documentclass[aps,prc,twocolumn,tightenlines,superscriptaddress]{revtex4}
\usepackage{epsfig,rotating}
\usepackage{multirow}

\newcommand{\nuc}[2]{\hbox{$^{#1}$#2}}

\begin{document}
\title{Collectivity at $N=50$: \nuc{82}{Ge} and \nuc{84}{Se}}

\author{A.\ Gade}
   \affiliation{National Superconducting Cyclotron Laboratory,
      Michigan State University, East Lansing, Michigan 48824}
   \affiliation{Department of Physics and Astronomy,
      Michigan State University, East Lansing, Michigan 48824}
\author{T.\ Baugher}
    \affiliation{National Superconducting Cyclotron Laboratory,
      Michigan State University, East Lansing, Michigan 48824}
    \affiliation{Department of Physics and Astronomy,
      Michigan State University, East Lansing, Michigan 48824}
\author{D.\ Bazin}
    \affiliation{National Superconducting Cyclotron Laboratory,
      Michigan State University, East Lansing, Michigan 48824}
\author{B.\,A.\ Brown}
    \affiliation{National Superconducting Cyclotron Laboratory,
      Michigan State University, East Lansing, Michigan 48824}
    \affiliation{Department of Physics and Astronomy,
      Michigan State University, East Lansing, Michigan 48824}
\author{C.\,M.~Campbell}
    \affiliation{National Superconducting Cyclotron Laboratory,
      Michigan State University,
      East Lansing, Michigan 48824}
\author{T.\ Glasmacher}
    \affiliation{National Superconducting Cyclotron Laboratory,
      Michigan State University, East Lansing, Michigan 48824}
    \affiliation{Department of Physics and Astronomy,
      Michigan State University, East Lansing, Michigan 48824}
\author{G.\ F.\ Grinyer}
    \affiliation{National Superconducting Cyclotron Laboratory,
      Michigan State University, East Lansing, Michigan 48824}
\author{M.\ Honma}
    \affiliation{Center for Mathematical Sciences, University of Aizu, Tsuruga,
      Ikki-machi, Aizu-Wakamatsu, Fukushima 965-8580, Japan}
\author{S.\ McDaniel}
    \affiliation{National Superconducting Cyclotron Laboratory,
      Michigan State University, East Lansing, Michigan 48824}
    \affiliation{Department of Physics and Astronomy,
      Michigan State University, East Lansing, Michigan 48824}
\author{R.\ Meharchand}
    \affiliation{National Superconducting Cyclotron Laboratory,
      Michigan State University, East Lansing, Michigan 48824}
    \affiliation{Department of Physics and Astronomy,
      Michigan State University, East Lansing, Michigan 48824}
\author{T.~Otsuka}
    \affiliation{Department of Physics and Center for Nuclear Study,
     University of Tokyo, Hongo, Tokyo 113-0033, Japan}
    \affiliation{RIKEN, Hirosawa, Wako-shi, Saitama 351-0198, Japan}
\author{A.\ Ratkiewicz}
    \affiliation{National Superconducting Cyclotron Laboratory,
      Michigan State University, East Lansing, Michigan 48824}
    \affiliation{Department of Physics and Astronomy,
      Michigan State University, East Lansing, Michigan 48824}
\author{J.\,A.\ Tostevin}
    \affiliation{Department of Physics, Faculty of Engineering and
      Physical Sciences, University of Surrey, Guildford, Surrey GU2 7XH,
      United Kingdom}
\author{K.\ A.\ Walsh}
    \affiliation{National Superconducting Cyclotron Laboratory,
      Michigan State University, East Lansing, Michigan 48824}
    \affiliation{Department of Physics and Astronomy,
      Michigan State University, East Lansing, Michigan 48824}
\author{D.\ Weisshaar}
    \affiliation{National Superconducting Cyclotron Laboratory,
      Michigan State University, East Lansing, Michigan 48824}
\date{\today}

\begin{abstract}
The neutron-rich $N=50$ isotones \nuc{82}{Ge} and \nuc{84}{Se} were investigated
using intermediate-energy Coulomb excitation on a \nuc{197}{Au} target and
inelastic scattering on \nuc{9}{Be}. As typical for intermediate-energy Coulomb
excitation with projectile energies exceeding 70~MeV/nucleon, only the first
$2^+$ states were excited in \nuc{82}{Ge} and \nuc{84}{Se}. However,
in the inelastic scattering on a \nuc{9}{Be} target, a strong population of the
first $4^+$ state was observed for \nuc{84}{Se}, while there is no indication of
a similarly strong excitation of the corresponding state in the neighboring
even-even isotone \nuc{82}{Ge}. The results are discussed in the framework of
systematics and shell-model calculations using three different effective
interactions.
\end{abstract}

\maketitle
\section{Introduction}

The selenium and germanium isotopic chains exhibit a complex nuclear structure
and have long been a rich testing ground for nuclear structure models. Their
properties are driven by shape coexistence and rapid shape changes all the way
from the $N=Z$ line into the
$A\approx 70$ mass
region~\cite{Sug03,Kot90,Toh00,Toh01,For87,Sta07,Obertelli09,Fis00,Ham74,Ham76,Hee86,Hur07,Lju08}.
On the other side of the nuclear chart, the most
neutron-rich selenium and germanium isotopes accessible for experiments are
around the magic neutron number $N=50$. Considerable experimental and
theoretical efforts have recently been focused in this region  on the investigation of the shell
structure approaching the doubly-magic nucleus
\nuc{78}{Ni}, see for
example~\cite{Verney2007,Lebois2009,Zhang2004,Walle2007,Hakala2008}.
The description of nuclei in this region poses a challenge for shell-model
calculations since the full $pf$ shell and the neutron $g_{9/2}$ intruder
orbital would be needed with the corresponding effective interaction. Presently,
smaller configuration spaces have to be used, typically starting from a
\nuc{56}{Ni} core and including the $p_{3/2}$, $f_{5/2}$, $p_{1/2}$
and $g_{9/2}$ orbitals~\cite{Lisetskiy2004,Honma2009}. Experimental information
is important to guide the emerging shell-model effective interactions in
this region.

In the present paper we report the experimental results of the
intermediate-energy projectile Coulomb excitation and inelastic scattering on
a \nuc{9}{Be} target for the $N=50$ isotones \nuc{82}{Ge} and
\nuc{84}{Se}. While Coulomb excitation with fast projectile beams allows
for the sensitive study of the $B(E2; 0^+_{gs} \rightarrow 2^+_1) \equiv B(E2
\uparrow)$ excitation strength in even-even nuclei -- a measure of the low-lying
quadrupole collectivity -- \nuc{9}{Be}-induced inelastic scattering provides
access to collective structures beyond the first $2^+$ state. Measured $B(E2
\uparrow)$ electric quadrupole excitation strengths along line of $N=50$
isotones are compared to large-scale shell-model calculations with three
different effective interactions. The evolution of collectivity along the Se and
Ge chains is further confronted with mean-field calculations. The
population of higher-lying states in the \nuc{9}{Be}-induced inelastic
scattering of \nuc{84}{Se} is discussed in comparison
to inelastic proton and $\alpha$ scattering on stable selenium isotopes.

\section{Experiment}
The measurements were performed at the National Superconducting Cyclotron
Laboratory (NSCL) on the campus of Michigan State University. The neutron-rich
projectile beams containing \nuc{82}{Ge} and \nuc{84}{Se} were produced
in-flight by fragmentation of a 140~MeV/u \nuc{86}{Kr} primary
beam provided by the Coupled-Cyclotron Facility at NSCL. \nuc{9}{Be} foils with
thicknesses
of 432~mg/cm$^2$ and 329~mg/cm$^2$, respectively, served as production
targets for the two different secondary beams. The fragments of interest
were selected with the A1900 fragment separator~\cite{a1900}; an achromatic
210~mg/cm$^2$ aluminum wedge degrader located at the mid-acceptance position of
the fragment separator was used. The total momentum acceptance was restricted to 2\%
for \nuc{82}{Ge} and 1\% for \nuc{84}{Se}. The setting optimized on \nuc{84}{Se}
resulted in a pure ($> 99.5\%$) secondary beam. The purity of the cocktail beam
containing \nuc{82}{Ge} was 32\%.

\begin{figure}[h]
        \epsfxsize 7.8cm
        \epsfbox{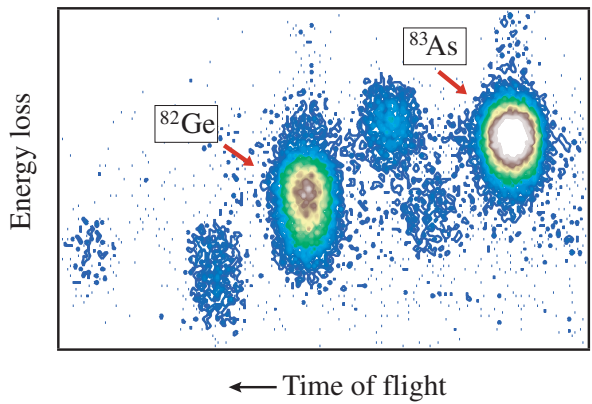}
\caption{\label{fig:pid} (Color online) Particle identification for the cocktail
  beam optimized on \nuc{82}{Ge} after interaction with the gold target. Plotted
  is the energy loss measured with the S800
  ionization chamber versus the time of flight measured between the plastic
  trigger
  scintillator at the back of the S800 focal plane and a timing scintillator
  at the spectrograph's object position.}
\end{figure}

Gold and beryllium targets used to induce projectile Coulomb
excitation
and inelastic scattering, respectively, were located at the target position of
the S800 spectrograph~\cite{s800}. The identification of the scattered
projectiles and the
trajectory reconstruction used to derive the scattering angles on an
event-by-event basis utilized the detection systems of the
spectrograph's focal plane, consisting of an ionization chamber, two
xy-position-sensitive cathode-readout drift chambers and a plastic timing
scintillator~\cite{Yurkon1999}. An example of the identification of the
scattered projectiles emerging from the gold target is shown in
Fig.~\ref{fig:pid} for the setting optimized on \nuc{82}{Ge}, where the
energy loss measured with the S800 ionization chamber versus the ion's time of
flight measured between two plastic scintillators is displayed. \nuc{82}{Ge} can be clearly
separated from the other constituents of the cocktail beam that contained
\nuc{83}{As} as the largest contaminant.

The reaction target located in front of the S800 spectrograph was surrounded by
the high-resolution
$\gamma$-ray detection system SeGA, an array of 32-fold segmented high-purity
germanium detectors~\cite{sega}. The segmentation of
the detectors allowed for an event-by-event Doppler reconstruction.
The angle of the $\gamma$-ray emission was deduced from the
position of the segment that registered the highest energy
deposition. The detectors were arranged in two rings
(90$^{\circ}$ and 37$^{\circ}$ central angles with respect to the beam
axis). The 37$^{\circ}$ ring was
equipped with seven detectors, while ten and nine detectors were located at
90$^{\circ}$ for the Coulomb excitation and \nuc{9}{Be}-induced inelastic scattering
measurements, respectively. The energy-dependent photopeak efficiency of the
setups was determined with standard \nuc{152}{Eu} and \nuc{226}{Ra} calibration
sources.

\section{Results and Discussion}
\subsection{Intermediate-energy Coulomb excitation}

Coulomb excitation is a widely used experimental technique to assess
the low-lying quadrupole collectivity in nuclei. In projectile Coulomb
excitation, exotic nuclei, produced as beams of ions, are scattered off
stable high-$Z$ targets and are detected in coincidence with the de-excitation
$\gamma$ rays that tag and quantify the inelastic process~\cite{Gla98,Mot95,Wan97}. While
beam energies
below the Coulomb barrier prevent nuclear contributions to the excitation
process, very peripheral collisions must be chosen in the regime of
intermediate-energy Coulomb scattering to exclude nuclear contributions. This
can be realized by restricting the data
analysis to scattering events at very forward angles, corresponding
to large minimum impact parameters, $b_{min}$, in the collisions
of projectile and target nuclei~\cite{Gla98}. Impact parameters exceeding
$1.2(A_p^{1/3} + A_t^{1/3})+2~$fm (``touching sphere + 2 fm'') have been proven
sufficient to ensure the dominance of the electromagnetic
interaction~\cite{Gade08a,Delaunay2007,cook06}.

\begin{figure}[h]
        \epsfxsize 6.0cm
        \epsfbox{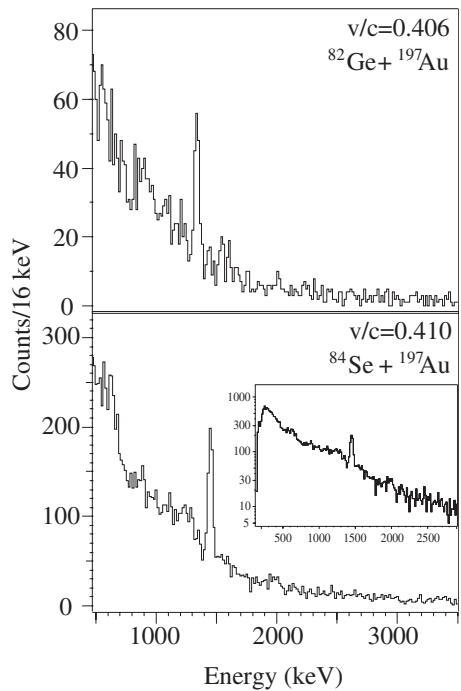}
\caption{\label{fig:gamma} Projectile Coulomb excitation of
  \nuc{82}{Ge} (upper panel) and \nuc{84}{Se} (lower panel). The $\gamma$-ray
  energies are event-by-event Doppler reconstructed into the rest frame of the
  projectile using the angle information obtained from the segmentation of the
  SeGA detectors. Only one $\gamma$-ray transition, the de-excitation of the
  $2^+_1$ state, was observed in each nucleus. The inset shows a wider range of
the \nuc{84}{Se} $\gamma$-ray spectrum on a logarithmic scale; no other
transitions were observed.}
\end{figure}

\begin{figure}[h]
        \epsfxsize 6.2cm
        \epsfbox{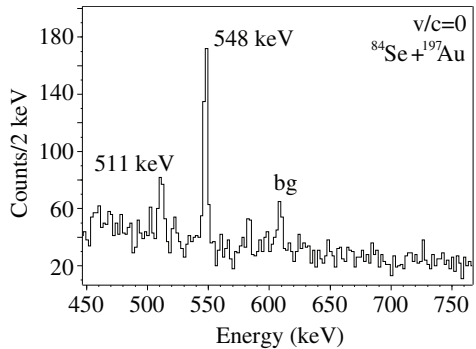}
\caption{\label{fig:Au} Coulomb excitation of \nuc{197}{Au} induced by the
  \nuc{84}{Se} projectile beam passing through the gold target (laboratory
  frame, $v/c=0$). The $\gamma$-ray
  transition corresponds to the de-excitation
  of the 547.5~keV $7/2^+$ state to the $3/2^+$ ground state.}
\end{figure}

In the present work, gold targets of thicknesses
256~mg/cm$^2$ and 184~mg/cm$^2$ for \nuc{82}{Ge} and \nuc{84}{Se},
respectively, were used to induce the Coulomb excitation. The
mid-target energies of the \nuc{82}{Ge} and \nuc{84}{Se} beams were
89.4~MeV/nucleon and 95.4~MeV/nucleon, respectively, resulting in a minimum
impact parameter of $b_{min} = 14.2$~fm for both the \nuc{82}{Ge}+\nuc{197}{Au}
and \nuc{84}{Se}+\nuc{197}{Au} collisions. Correspondingly, maximum scattering
angles in the laboratory system of $\theta_{max}=2.05^{\circ}$ and $1.99^{\circ}$
were chosen for the analysis of \nuc{82}{Ge} and \nuc{84}{Se}, respectively. The
target Coulomb excitation of the first excited $7/2^+$ state
in \nuc{197}{Au} by the electromagnetic field of the projectiles passing through
the target was observed. Figures \ref{fig:gamma} and \ref{fig:Au} show the
$\gamma$-ray spectra detected in coincidence with the different scattered
projectiles.

Angle-integrated Coulomb excitation cross sections $\sigma(\theta \leq
\theta_{max})$ were determined from the efficiency-corrected $\gamma$-ray
intensities of the $2^+_1 \rightarrow 0^+_{gs}$ transitions relative to the
number densities of the gold targets and the number of projectiles passing
through
the targets. The efficiencies were corrected for the Lorentz boost and the
$\gamma$-ray angular distribution in intermediate-energy Coulomb
excitation~\cite{Win79} and absorption in the gold target. The
semi-classical Winther-Alder theory~\cite{Win79} was used to derive $B(E2
\uparrow)$
excitation strengths from the angle-integrated cross sections. To test the setup
and analysis procedures, the $B(E2; 3/2^+ \rightarrow 7/2^+)$ electromagnetic
transition strength in \nuc{197}{Au} was determined from the Coulomb excitation
of the \nuc{197}{Au} target induced by the \nuc{82}{Ge}, \nuc{83}{As} and
\nuc{84}{Se} projectiles. Table~\ref{tab:exp_c} summarizes the angle-integrated
Coulomb excitation cross sections and extracted $B(E2 \uparrow)$ values. The
results
for \nuc{82}{Ge} and \nuc{197}{Au} from this work agree with the literature
values~\cite{Padilla2005,Zhou95}.

\begin{table}[h]
\caption{Experimental results for \nuc{82}{Ge} and \nuc{84}{Se}. The mean
  lifetimes, $\tau$, are deduced from the $B(E2\uparrow)$ strengths. The Coulomb
  excitation of the gold target by \nuc{82}{Ge}, \nuc{84}{Se} and \nuc{83}{As}
  projectiles was quantified as a cross check of the experimental setup and
  analysis procedures. If available, the literature values are given.}
\begin{ruledtabular}
\begin{tabular}{l|cc|cc}
  & \nuc{82}{Ge}& Ref.~\cite{Padilla2005} &\nuc{84}{Se}& \\
\hline
$E(2^+_1)$ (keV) & 1348&1348 &1454&1454 \\
$\sigma(\theta \leq \theta_{max})$ (mb)&258(36) & & 199(22)&  \\
$B(E2 \uparrow)$~$(e^2$b$^2)$&0.128(22) & 0.115(20) & 0.105(15)& -\\
$\tau$ (ps) & 0.72(12)& 0.80(14) &0.60(9)& -\\
\hline
\nuc{197}{Au} & &Ref.~\cite{Zhou95}  & &Ref.~\cite{Zhou95} \\
\hline
$\sigma(\theta \leq \theta_{max})$ (mb)&152(28) & & 150(16)&  \\
$B(E2 \uparrow)$~$(e^2$b$^2)$& 0.476(94)& 0.449(41) &0.441(64)&  0.449(41)\\
                              & 0.424(76)\footnote{From the excitation of the
                                \nuc{197}{Au} target by \nuc{83}{As} in the
                                cocktail beam that contained \nuc{82}{Ge}} &  &       &
\label{tab:exp_c}
\end{tabular}
\end{ruledtabular}
\end{table}

\begin{figure}[h]
        \epsfxsize 8.4cm
        \epsfbox{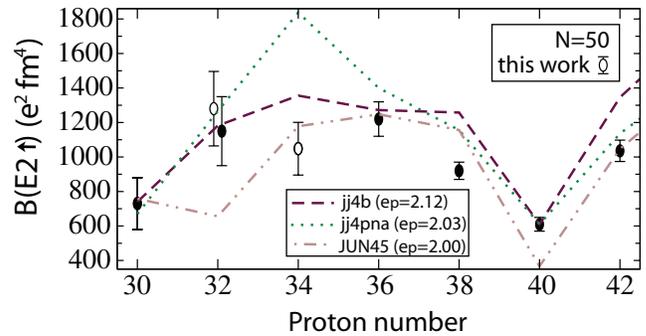}
\caption{\label{fig:shell} (Color online) $B(E2;0^+ \rightarrow 2^+_1)$ systematics of the
  $N=50$ isotones. The evolution of quadrupole collectivity along $N=50$ is
  compared to
  shell-model calculations with the jj4b~\cite{Verney2007} ($e_p=2.12$),
  jj4pna~\cite{Verney2007} ($e_p= 2.03$) and the JUN45~\cite{Honma2009} ($e_p=2.00$) effective
  interactions. The effective proton charge, $e_p$, was chosen for each
  interaction to get closest to the experimental data. The need for fairly
  high effective charges of $e_p \approx 2$ illustrates the importance of core
  excitations across the $N=50$ shell gap which are outside of the employed
  model space. The experimental data stems from the present work and
  references~\cite{Padilla2005,Walle2007,Raman01,ensdefMo,Sta07,Obertelli09}.
}
\end{figure}

Figure~\ref{fig:shell} shows the systematics of the reduced electric quadrupole
excitation strength $B(E2;0^+ \rightarrow 2^+_1)$ for the $N=50$ isotones from
zinc to molybdenum. The experimental results are compared to shell-model
calculations using the jj4b, jj4pna~\cite{Verney2007} and the
JUN45~\cite{Honma2009} effective interactions. Similar to the work on $E2$
transition rates in $N=50$ isotones by Ji and
Wildenthal~\cite{Ji88}, proton effective charges of
$e_p\approx 2$ were used~\footnote{We note that the neutron effective charge is
  irrelevant since a closed neutron shell is assumed in the shell model and thus
  $A_n=0$ in $B(E2 \uparrow)= (A_p e_p+ A_n e_n)^2$}. The
need for a rather large proton effective charge, compared to $e_p=1.5$ typical
for calculations in the $sd$ shell, for
example, is indicative of missing neutron core excitations across the $N=50$
shell gap in the $f_{5/2},p_{3/2},p_{1/2},g_{9/2}$ model space. The three effective interactions differ markedly for \nuc{82}{Ge} and \nuc{84}{Se} while
they show very similar trends at $Z=30$ and for $Z \geq 36$. Calculations with
jj4b and jj4pna agree with each other and the experimental value for
\nuc{82}{Ge} while all three interactions differ at \nuc{84}{Se}, with JUN45
describing the excitation strength best.

It is apparent from the comparison in Fig.~\ref{fig:shell} that for the $N=50$
isotones the calculated $B(E2 \uparrow)$ values of \nuc{82}{Ge} and \nuc{84}{Se}
are particularly sensitive to details of the shell-model effective
interaction. The two valence proton orbitals being filled between $Z=28$ and
$Z=38$ are $f_{5/2}$ and $p_{3/2}$. The sensitivity arises from the details
on how these orbitals are occupied. Their filling is largely determined by their
effective single-particle energy (ESPE) gap and how it changes between
\nuc{78}{Ni} and \nuc{88}{Sr}. The ESPE were calculated for proton particle
states relative to a \nuc{78}{Ni} core and for proton hole states relative to
the proton configuration $(f_{5/2})^6 (p_{3/2})^4$ (the \nuc{88}{Sr} core). The size of the gap,
$\epsilon (p_{3/2})-\epsilon(f_{5/2})$, for the three Hamiltonians is given in
Table~\ref{tab:sm} along with the ground-state occupancy of the $f_{5/2}$ orbit,
$n(f_{5/2})$, in \nuc{82}{Ge} and \nuc{84}{Se}, respectively. In all three
cases, the $f_{5/2}$ orbital ESPE lies below that of the $p_{3/2}$.

\begin{table}[h]
\begin{center}
 \vspace{0.5cm}
\caption{ Effective single-particle energy (ESPE) gap between $f_{5/2}$ and $p_{3/2}$
  and the ground state occupancies for \nuc{84}{Se} and \nuc{82}{Ge} for the
  jj4pna, jj4b and JUN45 Hamiltonians.}
\begin{ruledtabular}
\begin{tabular}{ccccc}
&\multicolumn{2}{c}{$\epsilon (p_{3/2})- \epsilon (f_{5/2})$ (MeV)}&\multicolumn{2}{c}{$n(f_{5/2})$}\\
   & \nuc{78}{Ni} &  \nuc{88}{Sr} &  \nuc{84}{Se}$_{gs}$ &
   \nuc{82}{Ge}$_{gs}$ \\
\hline
jj4pna & 1.50 & 0.47 & 3.81 & 3.17\\
jj4b  & 0.39 & 0.72 & 4.00 & 2.84 \\
JUN45 & 0.97 & 1.11 & 4.40 & 3.16
\label{tab:sm}
\end{tabular}
\end{ruledtabular}
\end{center}
\end{table}

In the extreme case of a large gap, the \nuc{84}{Se} ground state has an
$(f_{5/2})^6$ closed-shell configuration
with an occupancy of $n(f_{5/2})=6$. For \nuc{84}{Se}, the $f_{5/2}$ occupancies
are correlated with the ESPE gap in \nuc{88}{Sr}. The most highly mixed
configuration is obtained with jj4pna and this is associated with a low energy
for the excited $2^+_1$ (1.18 MeV) together with a $B(E2 \uparrow)$ strength, which is
almost twice as large as experiment (see Fig.~\ref{fig:shell}). This is an
indication that the $p_{3/2} - f_{5/2}$ ESPE gap for this interaction is too
small at $Z=34$.

The ground state occupancy for \nuc{82}{Ge} on the other hand is correlated with
the ESPE gap for \nuc{78}{Ni}. The JUN45 Hamiltonian gives a $B(E2 \uparrow)$
value which is almost a factor of two smaller than experiment
(see Fig.~\ref{fig:shell}). However, for JUN45 there is considerable $E2$ strength
to the second $2^+$ state at 2.18 MeV (50\% of the strength to the 1.50 MeV
state). This fragmentation may be related to parts of the Hamiltonian that go
beyond the monopole terms that determine the ESPEs.

\begin{figure}[h]
        \epsfxsize 8.4cm
        \epsfbox{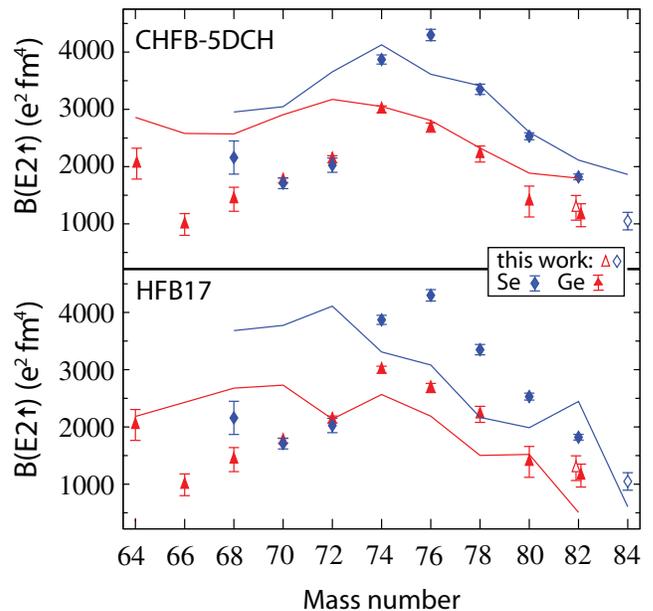}
\caption{\label{fig:HFB} (Color online) $B(E2;0^+ \rightarrow 2^+_1)$ systematics of the
  selenium and germanium isotopic chains compared to calculations with the
  constrained Hartree-Fock
  Bogoliubov with mapping on the 5-dimensional collective
  Hamiltonian~\cite{CHFB5DCH} approach (upper
  panel) and HFB calculations using the HFB-17
  parameterization~\cite{Goriely2009} (lower panel). Experimental values are
  taken from \cite{Raman01,Padilla2005,Obertelli09,Sta07,Lju08} and the present
  work.  }
\end{figure}

In Fig.~\ref{fig:HFB}, the $B(E2 \uparrow)$ values in the chains of selenium
($A=68-84$) and germanium
($A=64-82$) isotopes are compared to (beyond) mean-field calculations using the
constrained Hartree-Fock Bogoliubov (HFB) with mapping on the 5-dimensional collective
Hamiltonian~\cite{CHFB5DCH} approach that uses the D1S Gogny force, abbreviated
here by CHFB-5DCH, (upper panel) and to
results from HFB calculations using the HFB-17
parameterization~\cite{Goriely2009} (lower panel). For the comparison to
HFB-17, the quadrupole deformation parameters $\beta_2$ were translated into $B(E2
\uparrow)$ values via $\beta_2=\frac{4\pi}{3ZR^2}\sqrt{B(E2 \uparrow)/e^2}$,
where the sharp-surface radius $R$ was deduced consistently from
$R^2=5\langle r_c^2 \rangle/3-(0.88~fm)^2$ with $\langle r_c^2 \rangle^{1/2}$ the
root mean squared (rms) charge radius calculated within HFB-17 and 0.88~fm the
charge radius of the proton. CHFB-5DCH describes well the trend of the
quadrupole collectivity beyond $A=74$ but overpredicts the $B(E2)$ values
towards the $N=Z$ line where shape coexistence dominates the nuclear
structure. HFB-17 approximately reproduces the trend for the germanium isotopes
heavier than $A=70$ and at $N=Z$ but also overpredicts the collectivity for
$A=66-70$.  The selenium isotopes are not well described by HFB-17.

\subsection{Inelastic scattering from  \nuc{9}{Be}}

In addition to the Coulomb excitation, inelastic scattering off \nuc{9}{Be} was
measured. Here, one expects nuclear excitations to
dominate and to give access to states beyond the first $2^+$ excitation in
\nuc{84}{Se} and \nuc{82}{Ge}. A \nuc{9}{Be} target with a thickness of
188~mg/cm$^2$ was used to induce the inelastic excitations at 87.6~MeV/nucleon and
92~MeV/nucleon mid-target energies for \nuc{82}{Ge} and \nuc{84}{Se}, respectively.

\begin{figure}[h]
        \epsfxsize 6.8cm
        \epsfbox{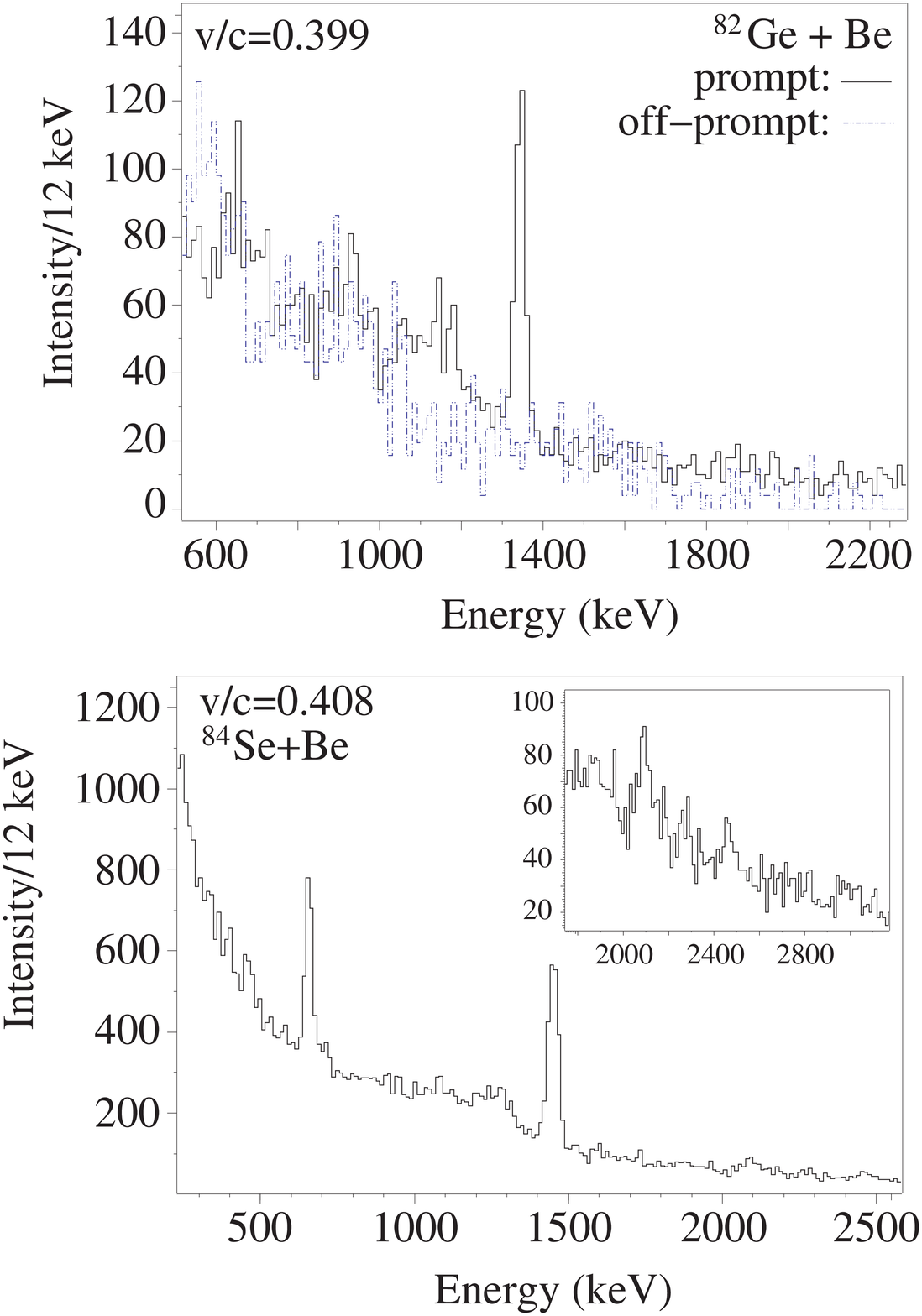}
\caption{\label{fig:GeBe} (Color online) Event-by-event Doppler-reconstructed $\gamma$-ray
  spectra from \nuc{82}{Ge}+\nuc{9}{Be} (upper panel) and \nuc{84}{Se} +
  \nuc{9}{Be} (lower panel) inelastic scattering. The only $\gamma$-ray
  transition in \nuc{82}{Ge} is the $2^+_1 \rightarrow 0^+$ transition. An
  in-beam background spectrum, obtained from an off-prompt SeGA time gate, is
  overlayed. In the \nuc{84}{Se} spectrum, an intense $\gamma$-ray transition at
  667(4)~keV is observed in addition to the decay of the $2^+_1$ state at
  1454~keV.
}
\end{figure}

The $\gamma$-ray spectra detected in coincidence with scattered \nuc{82}{Ge} and
\nuc{84}{Se} are shown in Fig.~\ref{fig:GeBe}. In the spectrum of \nuc{82}{Ge},
only the de-excitation of the first $2^+$ state at 1348~keV is
visible. Overlayed is the in-beam background obtained from an off-prompt gate on
the trigger-$\gamma$-timing. The only obvious structures above background in
\nuc{82}{Ge} are the full-energy peak of the $2^+_1 \rightarrow 0^+_1$
transition and its Compton edge. An excited state at 2287~keV in \nuc{82}{Ge}
has been tentatively assigned as the $4^+_1$ level from deep-inelastic
reactions and spectroscopy of \nuc{248}{Cm} fission
fragments~\cite{Angelis2007,Rzaca2007}. The corresponding $\gamma$-ray
transition energy for the decay to the $2^+_1$ level is
$E_{\gamma}=938$~keV. Unfortunately, the prompt
background is very high at this energy and it was only possible to establish an
upper limit of 60 counts in the full energy peak which corresponds to an upper
limit for the cross section of $\sigma(4^+_1) \leq 4.8$~mb for \nuc{82}{Ge}. The
cross sections
for \nuc{82}{Ge} are given in Table~\ref{tab:exp_Be}. The possible feeding from
the decay of the $4^+$ state is considered in the stated uncertainty of
$\sigma(2^+_1)$. We note that the low statistics for \nuc{82}{Ge} might obscure
the observation of additional weak feeding transitions.

In \nuc{84}{Se}, however, a second intense
$\gamma$-ray transition at 667~keV is clearly visible in addition to the $2^+_1
\rightarrow 0^+_1$ decay. As shown in the inset, there is also evidence for two
weaker $\gamma$-ray transitions at 2090~keV and 2462~keV. Excited states of
\nuc{84}{Se} are known from
\nuc{82}{Se}$(t,p)$\nuc{84}{Se}
two-neutron transfer reactions~\cite{Knight1974,Mullins1988}, from
$\gamma$-ray spectroscopy in \nuc{82}{Se}+\nuc{192}{Os} deep-inelastic
reactions~\cite{Zhang2004}, from prompt $\gamma$-ray detection following
fission-fragment spectroscopy~\cite{Prevost2004,Jones2006} and from $\beta$
decay~\cite{Kratz1975,Hoff1991}.

The intense transition at
667~keV was reported in several of the previous measurements and has been
attributed to the decay from the first $4^+$ state to the $2^+$ state, placing
the $4^+_1$ yrast level in \nuc{84}{Se} at 2122~keV excitation
energy~\cite{Zhang2004,Prevost2004,Jones2006}. The
transition at 2462~keV matches the energy of the ground-state decay of the
second $2^+$ state. With the $2^+_2$ state populated in the inelastic
scattering, one would also expect to see its 1007~keV transition to the first
$2^+$ state at a branching ratio of 42\% corresponding to about 80 counts. In
the spectrum there seems to be no clear indication of a peak with this
intensity, however, the background is high in this energy region. The
2090~keV transition could either be the
ground-state decay of the $(1^-)$ state at 2097(11)~keV reported only from
$(t,p)$ two-neutron transfer~\cite{Mullins1988}, or --  more likely -- the
2087~keV transition that depopulates an excited state at 3542~keV with suggested
$(2^+,3^-)$ assignment based on the population in $\beta$
decay~\cite{Kratz1975,Hoff1991}. The second, weaker (15.8(7)\%~\cite{nndc} ) decay branch of
this level to the $4^+$ would not have been visible in our spectrum. The cross
sections for \nuc{84}{Se} are summarized in Table~\ref{tab:exp_Be}. The feeding
from the decay of the $4^+$ state was taken into account for the determination of
the excitation cross section of
the $2^+$ state. The potential feeding by the 2090~keV transition and the
decay of the $2^+_2$ are included in the uncertainty.

\begin{table}[h]
\caption{Measured cross sections for $\nuc{82}{Ge}+\nuc{9}{Be}$ and
  $\nuc{84}{Se}+ \nuc{9}{Be}$. The $\sigma(2^+)$ for \nuc{84}{Se} is corrected
  for the feeding by the $4^+_1$ state. There is also
evidence for weak higher-energy transitions at 2090(10)~keV and
2462(11)~keV. The placement of the 2090~keV line in the level scheme is
unclear. The 2462~keV transition is likely the decay of the $2^+_2$ to the
ground state. The corresponding feeding by the $2^+_2 \rightarrow 2^+_1$
transition and the potential feeding by the 2090~keV transition has been taken
into account in the error bars for $\sigma(2^+_1)$. The potential feeding of the
$2^+_1$ state of \nuc{82}{Ge} by the $4^+_1$ decay has been folded into the
uncertainty of the $\sigma(2^+_1)$ cross section. }
\begin{ruledtabular}
\begin{tabular}{lcc}
$\sigma$ (mb)  & \nuc{82}{Ge}&\nuc{84}{Se}\\
\hline
$\sigma(2^+_1)$ & 27$^{+3}_{-6}$&20$^{+2}_{-6}$ \\
$\sigma(4^+_1)$ &$\leq$4.8 & 12.4(12) \\
$\sigma(2090$~keV) & -  & 5(1) \\
 $\sigma(2462$~keV) & - & 2.9(7)
\label{tab:exp_Be}
\end{tabular}
\end{ruledtabular}
\end{table}

\begin{figure}[h]
        \epsfxsize 6.0cm
        \epsfbox{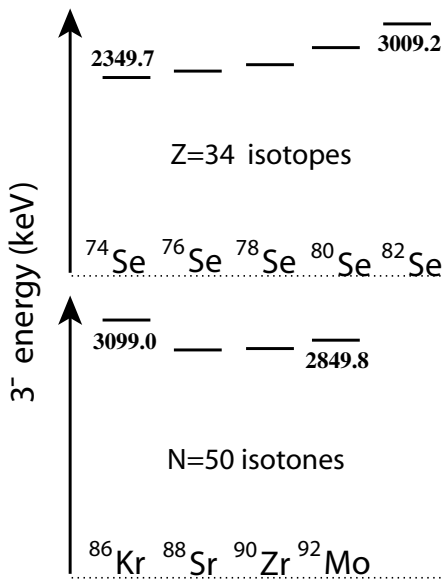}
\caption{\label{fig:3minus} Systematics of the $3^-_1$ states in the chain of Se
  isotopes (upper panel) and the $N=50$ isotones (lower panel). From both
  systematics one would expect the first $3^-$ state in \nuc{84}{Se} to lie
  above or around 3~MeV excitation energy.
}
\end{figure}

Inelastic $\alpha$~\cite{Ballester1988} and proton
scattering~\cite{Matsuki1983,Ogino1988,Delaroche1984} on the stable even-even
selenium isotopes \nuc{74-82}{Se} revealed that the $2^+_1$ states are excited
the strongest, followed by the first $3^-$ and, at markedly less cross section,
higher-lying $2^+$ states and the $4^+_1$ level. One might expect
\nuc{9}{Be}-induced
scattering to yield a similar population pattern, however, it seems for
\nuc{84}{Se} + \nuc{9}{Be} that the $4^+_1$ state is more strongly excited than
the $3^-$ state. Figure~\ref{fig:3minus} shows the systematics of $3^-_1$ states
in the selenium isotopes approaching \nuc{84}{Se} (upper panel) and in the $N=50$
isotone chain heavier than selenium (lower panel). From both systematics one
would expect the $3^-_1$ state in the $N=50$ selenium nucleus to be around or
above 3~MeV excitation energy. In fact, if \nuc{84}{Se} were to follow the trend
established by the lighter isotopes, the state at 3542~keV would emerge as a
good candidate. In the literature, $2^+$ and $(2^+,3^-)$  assignments based on
$\beta$-decay measurements~\cite{Kratz1975,Hoff1991,nndc} can be found for this
level. From the present work, both spin and parity assignments seem possible
although if this level turns out to be a high-lying $2^+$ state, the question
emerges where the
first $3^-$ state is located in \nuc{84}{Se} or why it is not as strongly
excited in the
inelastic scattering on \nuc{9}{Be} as one might expect from comparison
with inelastic proton and $\alpha$ scattering on the stable selenium
isotopes.

The role of the $4^+$ state in \nuc{84}{Se} and its strong population in the
\nuc{9}{Be}-induced inelastic
scattering emerge as interesting. From inelastic scattering of
polarized protons on \nuc{74-82}{Se}, Matsuki {\it et al.}~\cite{Matsuki1983}
present indications for a static or dynamic hexadecapole shape transition that
occurs between the light (\nuc{74,76,78}{Se}) and heavier (\nuc{80,82}{Se})
selenium isotopes and point out that the hexadecapole degree of freedom plays an
important role in the selenium isotopes. From inelastic proton scattering, Ogino
{\it et al.}~\cite{Ogino1988} find the hexadecapole strength fragmented strongly
for \nuc{74-82}{Se}. The transition strength to the $4^+_1$ state was found to
be rather weak except for the case of \nuc{82}{Se} where a transition strength of
1.3~spu was measured for the first $4^+$ state~\cite{Matsuki1983,Ogino1988}.

To help quantify these measured inelastic channel yields we have
performed macroscopic (deformed) coupled-channels calculations
\cite{Fresco}. The required projectile--\nuc{9}{Be} interactions were
estimated by double-folding the point neutron and proton densities
of \nuc{82}{Ge} and \nuc{84}{Se} (obtained from spherical Hartree-Fock
calculations~\cite{skx}) and of \nuc{9}{Be} (assumed a Gaussian with rms radius of
2.36~fm) with an effective nucleon-nucleon interaction \cite{Tos04}.
Radii $R_0 = 1.1 A^{1/3}$~fm were used in computing deformation
lengths, $B(E2 \uparrow)$, etc. For \nuc{82}{Ge}, the $B(E2 \uparrow)$
of Table \ref{tab:exp_c} is consistent with $\beta_2 = 0.2$, which gives a
calculated $\sigma(2_1^+) = 21.9$ mb in line with that measured.
Similarly for \nuc{84}{Se}, the $B(E2)$ of Table~\ref{tab:exp_c} corresponds to
$\beta_2 = 0.17$ ($\delta_2=0.83$ fm) giving $\sigma(2_1^+)= 15.2$~mb and, in
the absence of hexadecapole deformation $\sigma(4_1^+)= 0.05$~mb. When including
$|\beta_4| = 0.05$ (1.3~spu) as was deduced for the $^{82}$Se($4_1^+$) state
\cite{Matsuki1983}, $\sigma(4_1^+)= 1.12$~mb and with $|\beta_4| = 0.08$
(3~spu), being the maximum hexadecapole strength observed in the neighboring
selenium isotopes \cite{Ogino1988}, we obtain $\sigma(4_1^+)= 2.25$~mb. This
remains considerably adrift from the observed \nuc{84}{Se}($4_1^+$) yield. To
reproduce the measured cross sections of Table~\ref{tab:exp_Be},
using the coupled-channels model calculations described here,
would require the use of $\beta_2=\beta_4 \approx 0.2$, giving $\sigma(2_1^+) =
18.5$~mb and $\sigma(4_1^+) = 12.0$~mb.

\section{Summary}

In summary, the $B(E2;0^+_1 \rightarrow 2^+_1)$ excitation strengths were
measured for \nuc{82}{Ge} and \nuc{84}{Se} using intermediate-energy Coulomb
excitation. The quadrupole collectivity along the $N=50$ isotone chain from zinc
to molybdenum is compared to large-scale shell-model calculations with three
different effective interactions. The calculated $B(E2)$ values for
\nuc{82}{Ge} and \nuc{84}{Se} were found sensitive to the size of the ESPE gap
between the $p_{3/2}$ and $f_{5/2}$ orbits in \nuc{78}{Ni} and
\nuc{88}{Sr}. From comparison to experiment it is indicated that the relevant ESPE gap
for the jj4pna effective interaction is too small at $Z=34$ while the JUN45
Hamiltonian predicts the $E2$ strength fragmented over the first and
second $2^+$ states.

The quadrupole collectivity along the germanium and selenium chains is compared
to Skyrme Hartree-Fock Bogoliubov (HFB) calculations using the HFB-17 force and to
constrained HFB calculations extended by the generator coordinate method and
mapped onto a 5-dimensional collective quadrupole Hamiltonian (CHFB-5DCH with
Gogny D1S force). CHFB-5DCH describes well the trend of the
quadrupole collectivity beyond $A=74$ for both isotopic chains but overpredicts
the $B(E2 \uparrow)$ values towards the $N=Z$ line where shape-coexistence
occurs. HFB-17 approximately reproduces the trend for the germanium
isotopes  heavier than $A=70$ and at $N=Z$ but also overestimates the
collectivity between $A=66$ and 70. The selenium isotopes are not well described
by the HFB-17 parameterization.

In \nuc{9}{Be}-induced inelastic scattering, the first $4^+$ state of
\nuc{84}{Se} was populated with significant intensity while there was no
indication of a similarly strong population of the corresponding state in
\nuc{82}{Ge}. The excitation of the $4^+$ state is
discussed in comparison to inelastic $\alpha$ and proton scattering data on
stable selenium nuclei and coupled-channels calculations, however, its
explanation remains a challenge for future reaction theory and nuclear structure
calculations.

\begin{acknowledgments}
A.G. acknowledges discussions with Kirby W.\ Kemper (FSU),
Paul D.\ Cottle (FSU) and Lewis A.\ Riley (Ursinus). This work was supported by
the National Science Foundation under Grants No. PHY-0606007 and
PHY-0758099 and the UK Science and Technology Facilities
Council (Grant ST/F012012). A.G. is supported by the Alfred P. Sloan
Foundation.
\end{acknowledgments}


\end{document}